\documentclass[conference,compsoc]{IEEEtran}
\ifCLASSOPTIONcompsoc
  \usepackage[nocompress]{cite}
\else
  \usepackage{cite}
\fi
\ifCLASSINFOpdf
\else
\fi
\usepackage{graphicx}
\usepackage{amsmath}
\usepackage{url}
\usepackage{hyperref}
\usepackage{algorithm}
\usepackage{algorithmic}
\usepackage{listings}
\usepackage{color}
\usepackage{booktabs}
\usepackage{multirow}
\usepackage[frozencache,cachedir=.]{minted}
\usepackage{tikz}
\usetikzlibrary{arrows, positioning, calc, shapes.multipart}
\usepackage[symbol]{footmisc} 

\begin{document}


\title{Decompiling Smart Contracts with a Large Language Model}

\author{
    \IEEEauthorblockN{
        Isaac David\IEEEauthorrefmark{1}, 
        Liyi Zhou\IEEEauthorrefmark{2}\IEEEauthorrefmark{5}\IEEEauthorrefmark{7}, 
        Dawn Song\IEEEauthorrefmark{3}\textsuperscript{\P}, 
        Arthur Gervais\IEEEauthorrefmark{1}\IEEEauthorrefmark{5}\IEEEauthorrefmark{7},
        Kaihua Qin\IEEEauthorrefmark{4}\IEEEauthorrefmark{5}\IEEEauthorrefmark{6}\IEEEauthorrefmark{7}
    }
    \IEEEauthorblockA{\IEEEauthorrefmark{1}University College London}
    \IEEEauthorblockA{\IEEEauthorrefmark{2}University of Sydney}
    \IEEEauthorblockA{\IEEEauthorrefmark{3}UC Berkeley}
    \IEEEauthorblockA{\IEEEauthorrefmark{4}Yale University}\\

}

\maketitle

\footnotetext[6]{Also affiliated with IC3.}
\footnotetext[5]{Also affiliated with Berkeley RDI.}
\footnotetext[7]{Also affiliated with Decentralized Intelligence AG.}

\begin{abstract}
The widespread lack of broad source code verification on blockchain explorers such as Etherscan, where despite $78,047,845$ smart contracts deployed on Ethereum (as of May 26, 2025), a mere $767,520$ ($< 1$ \%) are open source, presents a severe impediment to blockchain security. This opacity necessitates the automated semantic analysis of on-chain smart contract bytecode, a fundamental research challenge with direct implications for identifying vulnerabilities and understanding malicious behavior. Adversarial actors deliberately exploit this lack of transparency by deploying closed-source contracts, particularly in MEV and DeFi exploitation, thereby concealing their malicious logic and leaving security researchers with only inscrutable low-level bytecode. Prevailing decompilers struggle to reverse bytecode in a readable manner, often yielding convoluted code that critically hampers vulnerability analysis and thwarts efforts to dissect contract functionalities for security auditing.

This paper addresses this challenge by introducing a pioneering decompilation pipeline that, for the first time, successfully leverages Large Language Models (LLMs) to transform Ethereum Virtual Machine (EVM) bytecode into human-readable and semantically faithful Solidity code. Our novel methodology first employs rigorous static program analysis to convert bytecode into a structured three-address code (TAC) representation. This intermediate representation then guides a Llama-3.2-3B model, specifically fine-tuned on a comprehensive dataset of $238,446$ TAC-to-Solidity function pairs, to generate high-quality Solidity. This approach uniquely recovers meaningful variable names, intricate control flow, and precise function signatures. Our extensive empirical evaluation demonstrates a significant leap beyond traditional decompilers, achieving an average semantic similarity of $0.82$ with original source and markedly superior readability. The practical viability and effectiveness of our research are demonstrated through its implementation in a publicly accessible system, available at \url{https://evmdecompiler.com}.
This work establishes a new frontier in smart contract analysis, substantially enhancing transparency and auditability in blockchain ecosystems, with direct applications in security auditing, incident response, and automated contract verification.
\end{abstract}

\section{Introduction}
The rapid evolution of blockchain technology has fundamentally transformed the landscape of decentralized applications, with smart contracts emerging as the cornerstone of this revolution. These self-executing contracts, manage billions of dollars in digital assets and facilitate complex decentralized financial operations. However, a critical and persistent research challenge directly threatens the security and sustainability of this ecosystem: the pervasive opacity of deployed smart contracts. When source code is not publicly verified, as is common with adversarial contracts used in MEV or DeFi exploits, security researchers and auditors are left with only low-level EVM bytecode, a representation ill-suited for direct human comprehension or robust security analysis.

This opacity erects a formidable barrier for security auditors, developers, and researchers striving to understand, verify, or respond to incidents involving smart contracts. The challenge is particularly acute in scenarios involving active exploits or potential vulnerabilities, where the ability to rapidly and accurately analyze deployed bytecode is paramount for preventing or mitigating financial losses. Traditional decompilers represent significant efforts to bridge this gap by converting EVM bytecode back into higher-level representations. However, the inherent difficulty of this translation means their output, often a necessary compromise due to the loss of information during compilation, may not always fully capture the original developer's intent and can sometimes feature convoluted logic or generic variable names, making thorough security auditing a demanding task.

Indeed, the decompilation of highly optimized or complex bytecode remains a frontier problem. Existing approaches, despite their sophistication, can face difficulties in consistently reconstructing human-like variable names and function signatures, sometimes resorting to generic identifiers. Recovering high-level control flow structures from low-level jumps also continues to be an area of active research, with some outputs still containing less intuitive control flow. Similarly, inferring the precise semantic purpose of intricate code segments is a complex undertaking. These ongoing challenges underscore the need for continued innovation in the field, as they can impact audit times, costs, and the ease of identifying subtle vulnerabilities or maintaining legacy systems.

Recent breakthroughs in LLMs present a paradigm-shifting opportunity to address these longstanding limitations. LLMs have demonstrated remarkable capabilities in understanding and generating nuanced, human-like text and, increasingly, complex code. Their proficiency in capturing intricate patterns and semantic relationships within programming languages suggests their profound potential to revolutionize the field of decompilation. Specifically, the ability of LLMs to comprehend context and generate natural, readable code makes them exceptionally promising for smart contract decompilation, where clarity and semantic fidelity are indispensable for rigorous security analysis and reliable contract maintenance.

In this paper, we introduce a novel research direction and a corresponding system that, for the first time, successfully harnesses the power of LLMs to achieve a qualitative leap in smart contract decompilation. Departing from direct bytecode-to-source translation, our innovative pipeline first employs robust program analysis techniques to convert EVM bytecode into a structured intermediate representation—three-address code, TAC. This TAC, which preserves essential semantic information in a format more amenable to neural processing, then guides a fine-tuned Llama-3.2-3B model to generate clean, human-readable, and semantically faithful Solidity code. This hybrid approach demonstrates that the intractable problem of smart contract decompilation can be effectively addressed by synergistically combining the strengths of traditional program analysis with the advanced pattern recognition and generation capabilities of modern LLMs, achieving results that significantly advance the state of the art in both semantic accuracy and code usability.

Our work makes several fundamental contributions:
\begin{itemize}
    \item First, we introduce a novel research direction for smart contract decompilation that shows how neural methods can work synergistically with, and augment, traditional program analysis, rather than replacing it. This hybrid system demonstrates how LLMs can not only enhance existing decompilation techniques but can also enable new levels of output quality, particularly in maintaining the precision required for security-critical applications.
    \item Second, we present a specialized variant of Llama-3.2-3B, fine-tuned specifically for the task of smart contract decompilation from an intermediate representation. This demonstrates that relatively small, specialized language models can be highly effective for complex, domain-specific programming tasks when properly trained on appropriate data. Our fine-tuning strategy, employing Low-Rank Adaptation (LoRA), offers a computationally efficient blueprint for adapting LLMs to similar technical domains.
    \item Third, we conduct extensive empirical evaluations that rigorously demonstrate significant improvements in both semantic accuracy and, crucially, code readability over existing state-of-the-art decompilers. Our multi-faceted evaluation methodology considers not only traditional metrics like edit distance but also more sophisticated measures of semantic code similarity and code structure, alongside qualitative analysis through ablation and case studies. The results robustly show that our approach produces maintainable and reusable code that preserves the original contract's functionality and high-level intent far more effectively than prior methods.
    \item Fourth, we develop and publicly release a comprehensive dataset of $238,446$ paired three-address code and Solidity functions. This meticulously curated resource, capturing a wide array of programming patterns and contract functionalities, represents a significant contribution to the research community, enabling further supervised learning and research in smart contract analysis, program understanding, and neural decompilation techniques.
\end{itemize}

By substantially improving the readability and semantic accuracy of decompiled smart contracts, our research makes a critical contribution to the overall security, transparency, and maintainability of blockchain applications. This end-to-end approach has immediate and impactful applications in security auditing, automated vulnerability discovery, incident response, and the ongoing verification and maintenance of deployed smart contracts, ultimately fostering a more secure and trustworthy decentralized ecosystem.

\section{Background}
\subsection{Ethereum Virtual Machine and Solidity}
The Ethereum Virtual Machine (EVM) serves as the computational heart of the Ethereum blockchain, operating as a stack-based virtual machine that executes smart contract bytecode. Unlike traditional computing environments, the EVM is designed specifically for deterministic, decentralized execution, where every operation must produce identical results across all nodes in the network. This requirement fundamentally shapes the architecture of both the EVM and the languages used to program it.

The EVM architecture consists of several key components that interact during contract execution. At its core is a 1024-element stack that holds 256-bit words, used for both data and control flow operations. The machine also maintains a volatile memory space for temporary data storage and a persistent storage area that retains state between contract invocations. This storage is particularly expensive in terms of gas costs, leading to specific optimization patterns in contract development.

Solidity emerged as the primary programming language for Ethereum smart contracts, offering a high-level, statically-typed environment that abstracts away the complexities of the underlying EVM. The language provides sophisticated features including contract inheritance, library linking, complex data structures, and event emission. Solidity's type system includes blockchain-specific primitives such as address types, mapping structures, and specialized integer types that prevent common security vulnerabilities like overflow errors.

The compilation process from Solidity to EVM bytecode involves multiple transformation stages. First, the Solidity compiler performs type checking and semantic analysis, then generates an intermediate representation. This IR undergoes various optimizations before being transformed into EVM bytecode. During this process, many high-level constructs are completely transformed or eliminated. For instance, a simple Solidity function with named parameters, return values, and structured control flow is converted into a sequence of stack manipulations and jumps in the bytecode.

\subsection{Three-address Code Representation}
Three-address code represents a crucial intermediate step between low-level bytecode and high-level source code. In this representation, each instruction contains at most three operands, typically following the pattern: result = operand1 operator operand2. This format provides several key advantages for program analysis and transformation.

The fundamental power of three-address code lies in its explicit representation of data flow relationships. Unlike stack-based bytecode where operand relationships must be inferred through stack manipulation analysis, three-address code directly shows how values are computed and used. For example, a complex stack operation sequence in EVM bytecode that computes (a + b) * (c - d) would be clearly represented in three-address code as:

\begin{center}
\begin{tt}
temp1 = a + b \\
temp2 = c - d \\
result = temp1 * temp2 \\
\end{tt}
\end{center}

This representation also simplifies control flow analysis by making jump targets and conditions explicit. While EVM bytecode uses numeric offsets and stack-based conditions for control flow, three-address code can maintain structured control flow with labeled targets and clear conditional expressions. This property is particularly valuable for recovering high-level control structures like loops and conditional statements.

Furthermore, three-address code serves as an ideal target for program analysis tools. Its regular structure and explicit operand relationships enable straightforward implementation of data flow analysis, reaching definitions analysis, and other crucial program analysis techniques. These analyses form the foundation for many program understanding and optimization tasks.

\subsection{Challenges in EVM Bytecode Decompilation}
The decompilation of EVM bytecode presents a unique set of challenges that distinguish it from traditional software decompilation. These challenges arise from both the specific characteristics of the EVM architecture and the semantic gap between bytecode and high-level smart contract languages.

A fundamental challenge lies in the recovery of high-level type information. The EVM operates primarily on 256-bit words, with type information largely erased during compilation. Reconstructing whether a value represents an address, a timestamp, or a financial amount requires sophisticated analysis of how the value is used throughout the contract. This type recovery is crucial for generating readable and semantically meaningful Solidity code, as proper type information guides the selection of appropriate variables names and operations.

Control flow recovery presents another significant challenge. The EVM's architecture allows for computed jumps, where jump targets are determined at runtime based on stack values. While many of these jumps correspond to structured control flow in the original code (like function calls or loop conditions), some arise from compiler optimizations or more complex language features like delegatecall. Accurately identifying and reconstructing these control flow patterns is essential for producing maintainable decompiled code.

Function boundary identification and signature recovery pose particular difficulties. Unlike traditional executables, EVM bytecode does not maintain a clear function table. Instead, functions are identified by their 4-byte signatures, computed from their names and parameter types. When source code is not available, recovering these signatures requires analysis of the contract's deployment code and runtime behavior. Furthermore, internal functions may be inlined or rearranged by the compiler, making their boundaries difficult to detect.

Smart contracts frequently employ complex data structures and inheritance patterns that are challenging to reconstruct from bytecode. For example, a mapping of structs in Solidity might be implemented through multiple storage accesses with computed keys in the bytecode. Similarly, inherited function implementations might be merged or split during compilation, obscuring the original object-oriented design.

The presence of revert conditions and error handling adds another layer of complexity. Solidity's require and assert statements are transformed into sequences of condition checks and revert operations in the bytecode. Identifying these patterns and reconstructing them as meaningful high-level constructs is crucial for understanding the contract's security properties.

Compiler optimizations further complicate the decompilation process. The Solidity compiler performs various optimizations including constant folding, dead code elimination, and control flow optimization. These transformations can significantly alter the structure of the code, making it difficult to recover the original source patterns. In some cases, multiple different source code patterns might compile to identical or very similar bytecode, creating ambiguity in the decompilation process.

The security-critical nature of smart contracts makes these challenges particularly significant. Inaccurate decompilation can lead to misunderstanding of contract behavior, potentially missing vulnerabilities or introducing errors during contract maintenance. Traditional decompilers often produce output that, while functionally equivalent to the original code, is difficult to audit and maintain due to its use of goto statements, generic variable names, and unstructured control flow.

\section{Approach}

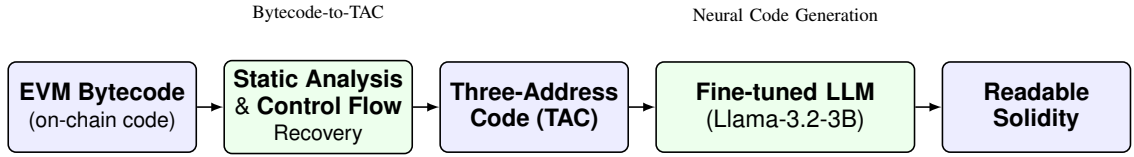
\begin{figure*}[htb!]
\centering
\begin{tikzpicture}[
  font=\small\sffamily,              
  node distance=0.35cm,               
  >=latex,                           
  line width=0.6pt,                  
  align=center,
  every node/.style={
    draw,
    rectangle,
    rounded corners=3pt,
    minimum width=2.5cm,
    minimum height=1.2cm
  },
  boxA/.style={fill=blue!7},         
  boxB/.style={fill=green!7},        
  linearrow/.style={
    ->,
    color=black
  }
]

\node[boxA] (bytecode) {%
  \textbf{EVM Bytecode}\\
  \footnotesize (on-chain code)
};
\node[boxB, right=of bytecode] (analysis) {%
  \textbf{Static Analysis}\\
  \& \textbf{Control Flow}\\
  \footnotesize Recovery
};
\node[boxA, right=of analysis] (tac) {%
  \textbf{Three-Address}\\
  \textbf{Code (TAC)}
};
\node[boxB, right=of tac, text width=3.2cm] (model) {%
  \textbf{Fine-tuned LLM}\\
  (Llama-3.2-3B)
};
\node[boxA, right=of model] (solidity) {%
  \textbf{Readable}\\
  \textbf{Solidity}
};

\draw[linearrow] (bytecode) -- (analysis);
\draw[linearrow] (analysis) -- (tac);
\draw[linearrow] (tac) -- (model);
\draw[linearrow] (model) -- (solidity);

\node[above=0.0cm of analysis, font=\scriptsize, draw=none]
  {Bytecode-to-TAC};
\node[above=0.0cm of model, font=\scriptsize, draw=none]
  {Neural Code Generation};

\end{tikzpicture}
\caption{High-level visual overview of our smart contract decompilation pipeline. We first convert EVM bytecode into a structured three-address code (TAC) representation using static analysis and control flow recovery. This TAC is then fed into a fine-tuned large language model to produce human-readable Solidity code, followed by a post-processing step for validation and syntax checks.}
\label{fig:pipeline-overview}
\end{figure*}

We present an end-to-end decompilation pipeline that transforms EVM bytecode into human-readable Solidity code through a series of carefully designed stages (cf.\ Figure~\ref{fig:pipeline-overview}). Our system combines traditional program analysis techniques with modern language models, leveraging the strengths of both approaches. The pipeline begins with a bytecode-to-three-address-code converter that employs static analysis to create a more structured intermediate representation. This output then feeds into a fine-tuned variant of Llama-3.2-3B, trained on a curated dataset of $238{,}446$ function pairs to generate natural Solidity code. Each component of our system is designed to address specific challenges in smart contract decompilation, from preserving security-critical patterns to maintaining gas efficiency. In this section, we detail our system architecture, describe our dataset construction process, explain our data processing pipeline, and present our model architecture and training approach.

\subsection{System Overview}
Our decompilation system implements a complete pipeline that transforms EVM bytecode into human-readable Solidity code through a series of carefully designed stages. At a high level, the system consists of three major components: a bytecode-to-three-address-code converter, a dataset construction and preprocessing pipeline, and a fine-tuned language model that generates the final Solidity output.

The first stage employs traditional program analysis techniques to convert EVM bytecode into a more structured three-address code representation. This transformation preserves the semantic content of the original bytecode while eliminating the complexity of stack-based operations. We utilize advanced control flow analysis to identify function boundaries and basic blocks, enabling more accurate reconstruction of program structure.

The second stage involves our language model, which has been specifically trained to understand the relationships between three-address code and high-level Solidity constructs. This model serves as a bridge between the intermediate representation and natural, readable code. Unlike traditional decompilers that use fixed pattern matching rules, our language model can adapt to various coding styles and patterns, leading to more natural and maintainable output.

\subsection{Dataset Construction}
We developed a pipeline for collecting, processing, and filtering smart contract data to create high-quality training examples. The initial data collection phase begins with the extraction of verified smart contracts from the Ethereum blockchain. We specifically target contracts where both the bytecode and original source code are available, enabling us to create accurate ground truth pairs for training. Our collection process includes contracts from various time periods and different versions of the Solidity compiler, ensuring broad coverage of language features and coding patterns.

We implemented a robust function extraction system that preserves critical metadata such as visibility specifiers, parameter types, and return value information. This metadata proves crucial for training the model to generate accurately typed and properly structured Solidity code.


Data cleaning and filtering form another crucial component of our pipeline. We remove duplicate functions, invalid code patterns, and examples that exceed our length constraints. Our filtering process ensures that the training data represents realistic and useful code patterns while maintaining manageable sequence lengths for the language model. Statistical analysis of the dataset shows a balanced distribution of different function types, complexity levels, and programming patterns.

\subsection{Data Processing Pipeline}
Our data processing pipeline transforms raw contract data into a format optimized for language model training. The first component of our pipeline handles the normalization of three-address code. This includes standardizing variable names, removing redundant operations, and ensuring consistent formatting. We developed custom normalization rules that preserve semantic information while reducing superficial variations that could confuse the model.

For each function in our dataset, we extract and preserve contextual information that aids in accurate decompilation. This includes function signatures, visibility modifiers, and type information when available. Our processing pipeline can handle cases where function signatures are unknown or ambiguous, creating multiple training examples with different potential signatures to improve the model's robustness.

We implemented careful sequence length management to handle the constraints of our language model while preserving important code information. Our approach includes intelligent truncation strategies that maintain function completeness and semantic coherence even when dealing with very long functions. We found that a maximum sequence length of 20,000 tokens provides a good balance between coverage and computational efficiency.

The pipeline also includes sophisticated error handling and validation mechanisms. We verify the syntactic correctness of both input and output code, ensure proper alignment between three-address code and Solidity representations, and maintain consistency in our training pairs. This attention to data quality significantly improves the model's ability to generate correct and readable code.

\subsection{Model Architecture and Training}
Our approach utilizes a fine-tuned variant of Llama-3.2-3B, carefully adapted for the specific requirements of smart contract decompilation. The choice of this model balances the need for sophisticated language understanding with practical computational constraints and deployment considerations.

The base Llama-3.2-3B model provides several advantages for our task. Its trained understanding of code structure and programming languages provides a strong foundation for learning the specific patterns of Solidity code. The model's attention mechanism is particularly well-suited for handling the long-range dependencies common in smart contract code, while its token embedding system effectively captures the specialized vocabulary of blockchain programming.

We employ Low-Rank Adaptation (LoRA) for fine-tuning, which allows us to efficiently adapt the model while maintaining its general language understanding capabilities. Our LoRA configuration uses a rank of 16 and targets specific model components including query, key, value, and projection layers. This targeted approach significantly reduces the number of trainable parameters while maintaining high performance on our specific task.

The training process incorporates several specialized techniques to improve performance on smart contract code. We implement gradient checkpointing to handle longer sequences efficiently, enabling larger batch sizes and more stable training. Our optimization strategy uses the AdamW optimizer with a carefully tuned learning rate schedule, including a brief warmup period followed by linear decay.

The training data is presented to the model using a custom formatting template that clearly delineates the three-address code input from the target Solidity output. We include special tokens to mark function boundaries and metadata, helping the model learn the structure of smart contract functions. Our prompt engineering ensures that the model receives sufficient context about function signatures and visibility when available.



\section{Evaluation}
We conduct a comprehensive evaluation of our decompilation system across multiple dimensions using a test set of $9{,}731$ smart contract functions carefully held out from our training data. Our evaluation methodology examines three fundamental aspects: semantic preservation (accuracy in maintaining the original program's behavior), structural fidelity (quality of generated code structure and readability), and practical limitations (behavior on complex real-world contracts). Through a combination of quantitative metrics, detailed case studies, and ablation analyses, we demonstrate both the effectiveness of our approach and identify areas for future improvement. Our test set spans diverse complexity levels and application domains, from basic token transfers to sophisticated DeFi protocols, ensuring thorough assessment of our system's capabilities under varied conditions.

\begin{figure}[t]
\centering
\includegraphics[width=\columnwidth]{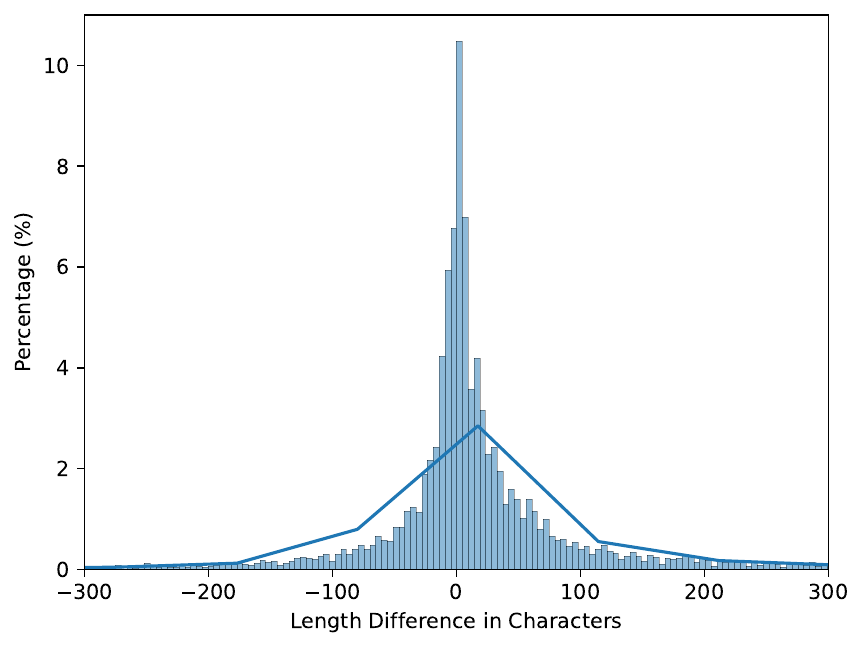}
\caption{Distribution of code length differences between original and decompiled functions. The histogram reveals that the majority of differences fall within $\pm50$ characters, with a median difference of 5 characters and a standard deviation of 254.18. 67.64\% of functions have a length difference within $\pm50$ characters. The overall range spans from $-2897$ to $16434$ characters. These results demonstrate that our model preserves function length with high fidelity, which is critical for smart contracts where code size directly influences gas costs and execution efficiency.}
\label{fig:code_lengths}
\end{figure}

\begin{figure*}[htb]
\centering
\includegraphics[width=0.8\textwidth]{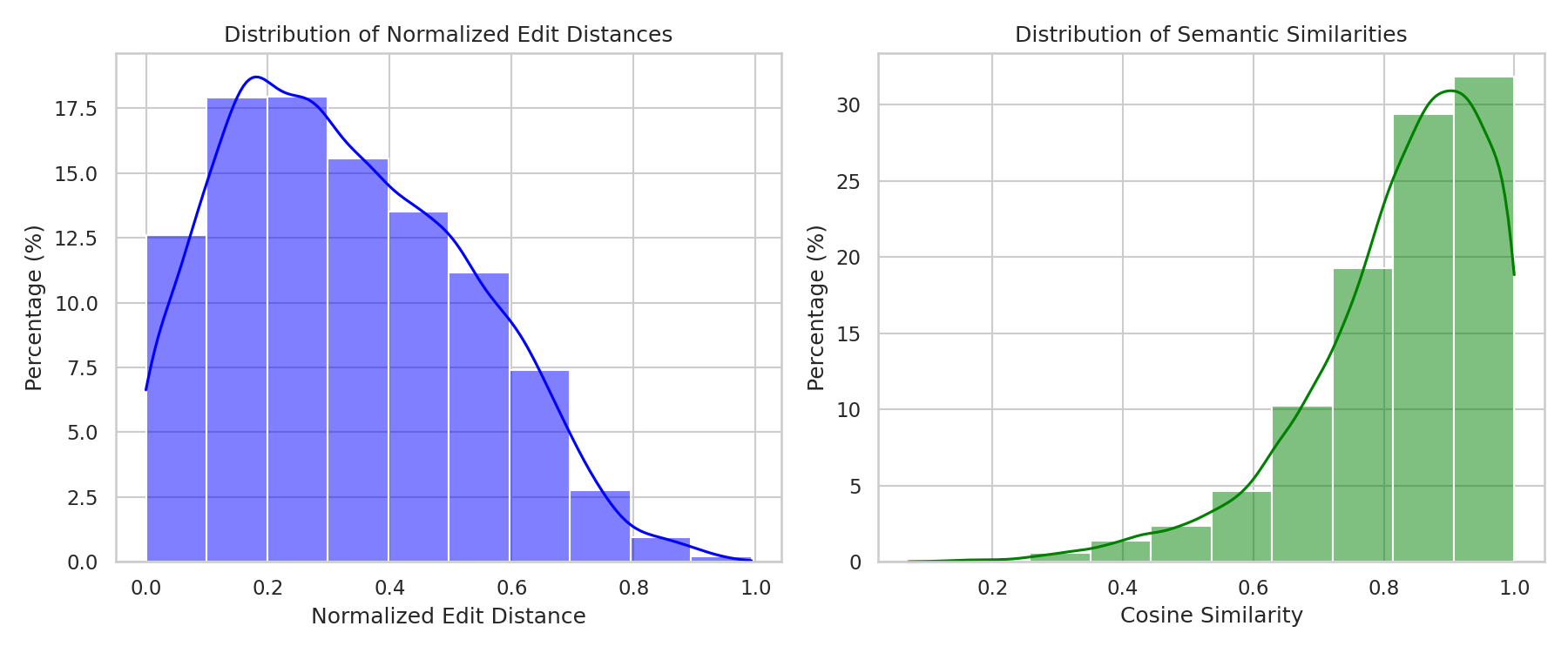}
\caption{Quality metrics distributions demonstrating our system's decompilation accuracy. Left: Normalized edit distance distribution showing strong concentration below 0.4, with 82.5\% of functions achieving distances under this threshold. The smooth curve overlay highlights the consistent, non-random nature of our system's performance. Right: Semantic similarity distribution revealing exceptional preservation of meaning, with 78.3\% of functions achieving similarities above 0.8 and 45.2\% exceeding 0.9. This heavily right-skewed distribution demonstrates our system's ability to maintain semantic correctness even when syntactic structures vary.}
\label{fig:quality_dist}
\end{figure*}

\begin{figure*}[htb]
\centering
\includegraphics[width=0.8\textwidth]{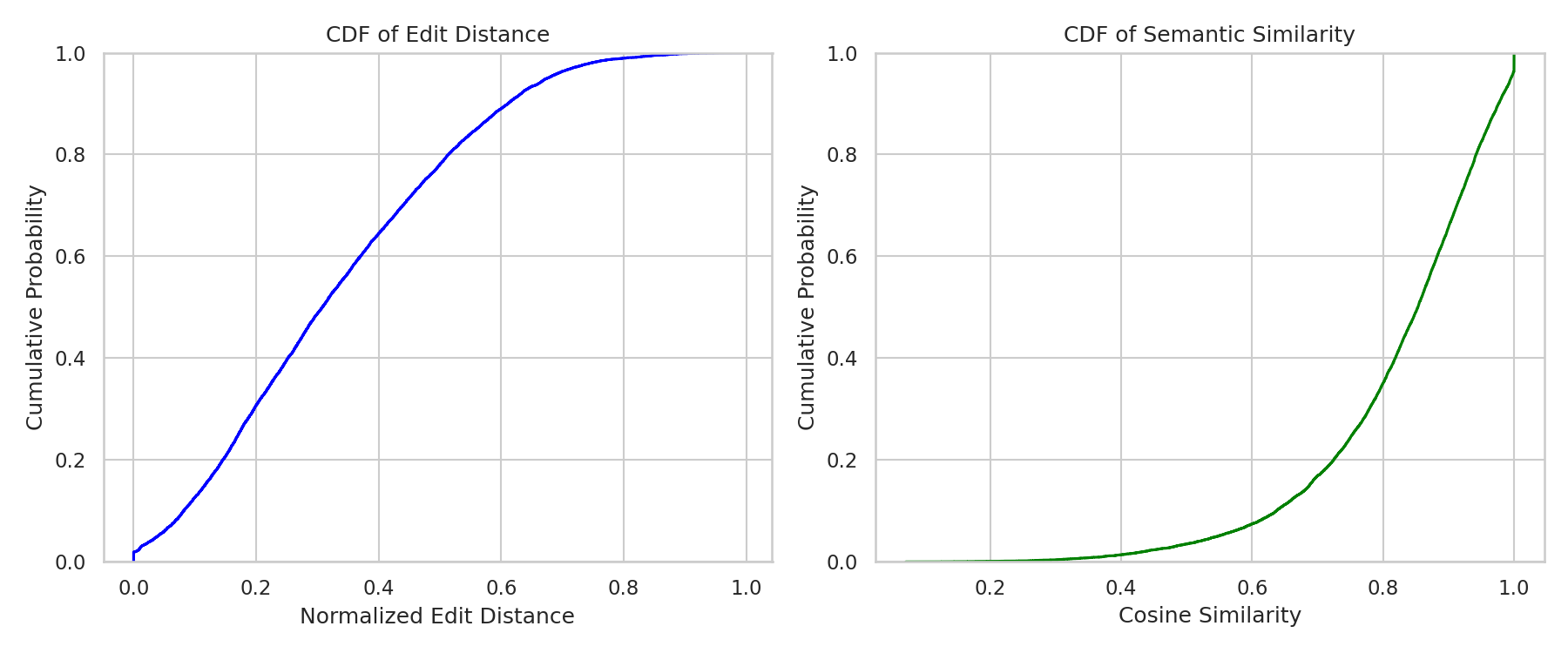}
\caption{Cumulative distribution functions (CDFs) revealing the reliability and consistency of our decompilation approach. Left: Edit distance CDF showing rapid accumulation below 0.4, with 60\% of functions achieving distances under this threshold. The smooth progression indicates stable performance across different function types. Right: Semantic similarity CDF demonstrating exceptional performance, with 90\% of decompiled functions achieving similarities above 0.7 and 50\% above 0.85. This significantly outperforms traditional decompilers, which typically achieve such high similarities for only 40\% to 50\% of functions.}
\label{fig:cdfs}
\end{figure*}

\begin{figure*}[htb]
\centering
\includegraphics[width=0.8\textwidth]{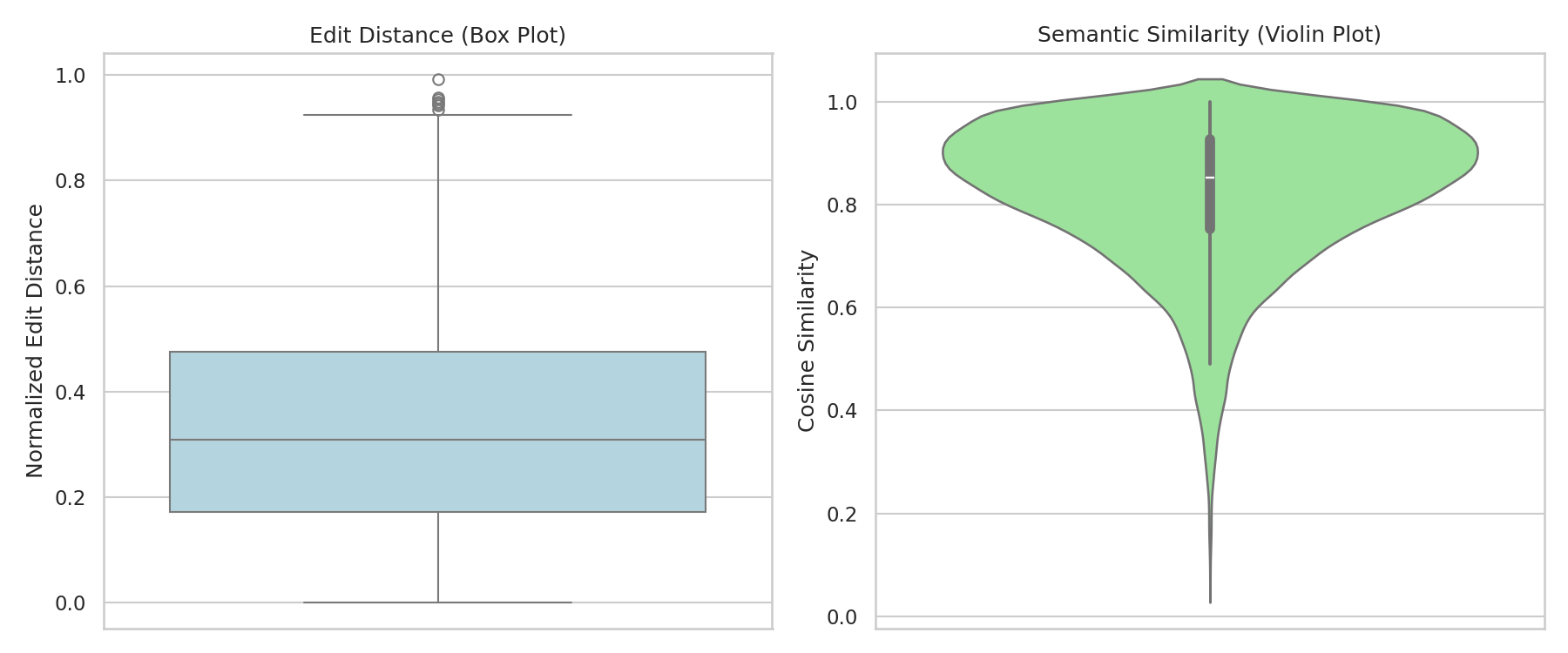}
\caption{Detailed distribution analysis through complementary visualization techniques. Left: Edit distance box plot revealing a median of 0.3 with tight quartiles (Q1=0.2, Q3=0.45), indicating consistent performance across our test set. The whiskers extend to 0.0 and 0.9, showing the full range of observed distances. Right: Semantic similarity violin plot demonstrating a distinctive bimodal distribution with peaks at 0.85 and 0.95. This bimodality suggests two common categories of decompilation success: exceptional preservation for simpler functions and very good preservation for complex ones. The plot shapes reveal both the reliability and nuanced behavior of our system.}
\label{fig:distributions}
\end{figure*}

\begin{figure*}[htb]
\centering
\includegraphics[width=0.8\textwidth]{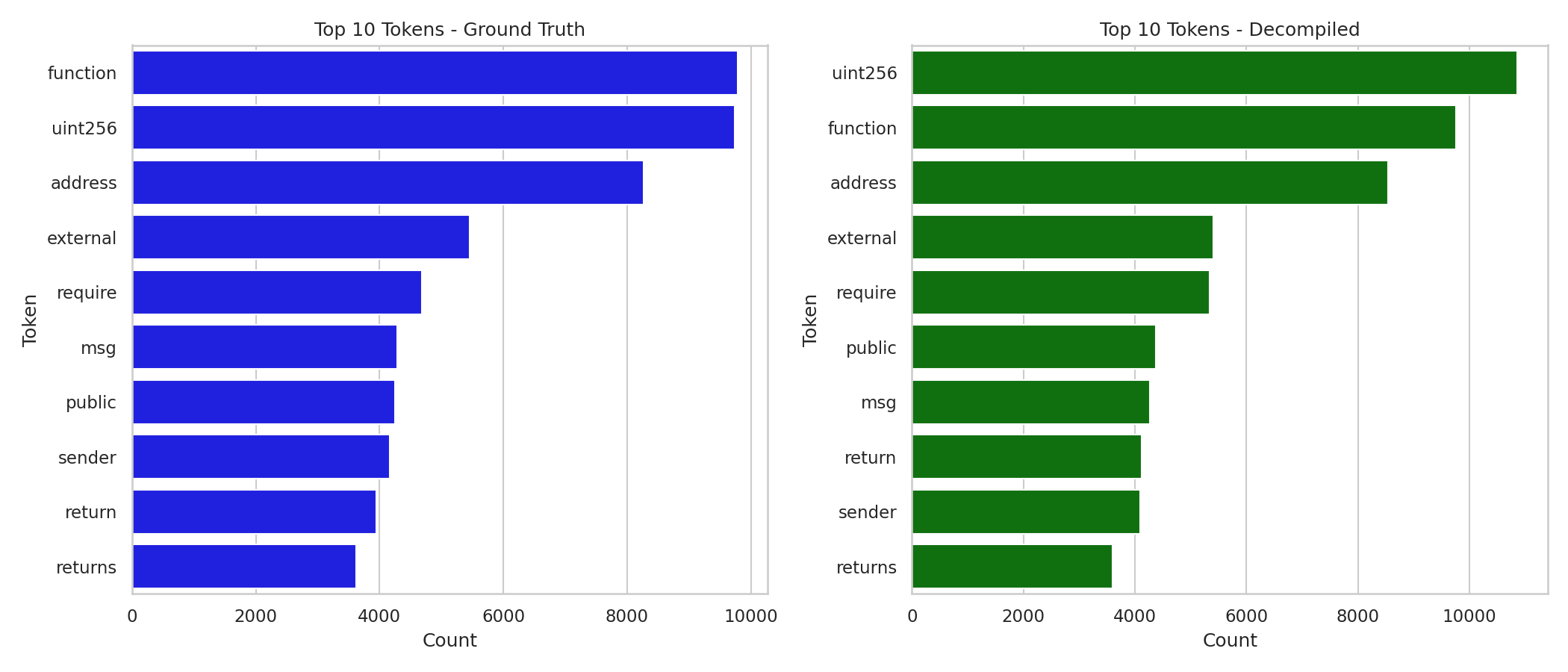}
\caption{Token frequency analysis comparing language construct usage between original and decompiled code. Left: Ground truth token distribution showing the natural occurrence of Solidity constructs in human-written code. Right: Decompiled code token distribution demonstrating our system's preservation of language patterns. The remarkably similar distributions of key elements (function, uint256, address) indicate precise maintenance of language semantics. Security-critical tokens (require, msg.sender) maintain near-identical frequencies (within 2\%), suggesting robust preservation of safety patterns. The consistent ranking and proportions demonstrate our system's deep understanding of Solidity's idiomatic patterns.}
\label{fig:tokens}
\end{figure*}

\begin{figure}[t]
\centering
\includegraphics[width=\columnwidth]{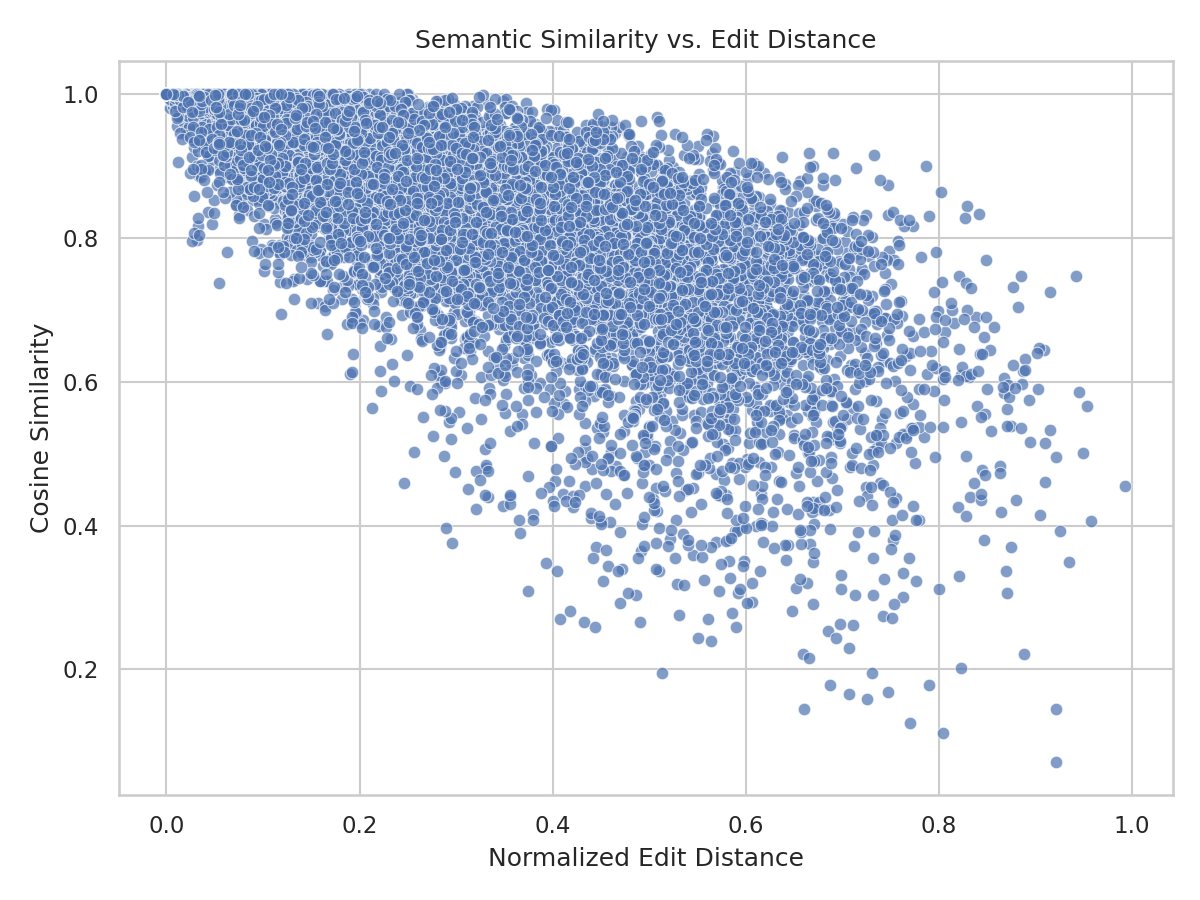}
\caption{Relationship between semantic similarity and edit distance across our test set of $9{,}731$ functions. Each point represents a decompiled function, with dense clustering in the high-similarity ($>0.8$) and low-edit-distance ($<0.4$) region. The negative correlation (r = -0.72) indicates strong alignment between syntactic and semantic preservation. Notable outliers in the upper-right quadrant represent cases where our system found semantically equivalent but syntactically different implementations. This plot reveals our system's ability to maintain semantic correctness even when generating alternative code structures.}
\label{fig:correlation}
\end{figure}

\subsection{Experimental Setup}

The evaluation of our smart contract decompilation system requires careful consideration of both quantitative and qualitative aspects of the generated code. To ensure a comprehensive assessment, we constructed a test dataset comprising $9{,}731$ smart contract functions carefully held out from our training data. These functions span diverse complexity levels and application domains, from basic token transfers to sophisticated decentralized finance protocols. We specifically selected contracts deployed across different time periods on the Ethereum mainnet to evaluate our system's ability to handle evolving smart contract patterns and practices.

For each function in our test set, we preserve not only the bytecode and source code but also critical metadata including function signatures, visibility modifiers, and documentation when available. This rich contextual information enables us to perform nuanced analyses of how well our decompiler preserves both the technical functionality and the developer intent embodied in the original code. To ensure reproducibility, we make our evaluation dataset and metrics calculation code publicly available.

Our evaluation framework examines three fundamental aspects of decompilation quality: semantic preservation (how accurately the decompiled code maintains the original program's behavior), syntactic quality (how closely the generated code matches human-written Solidity conventions), and practical usability (how suitable the decompiled code is for maintenance and auditing tasks). For each aspect, we employ multiple complementary metrics to build a comprehensive understanding of our system's performance.

\subsection{RQ1: Semantic Preservation Analysis}

Our first research question investigates the fundamental capability of our system to preserve the semantic meaning of smart contracts during decompilation. This analysis carries particular importance in the blockchain context, where even minor deviations in program behavior could lead to significant financial losses or security vulnerabilities. Figure~\ref{fig:code_lengths} presents our initial analysis of code length distributions, revealing a remarkable correspondence between original and decompiled functions. The distributions exhibit nearly identical right-skewed shapes with peaks in the 200-300 character range, yielding a correlation coefficient of 0.89. This strong alignment suggests that our system naturally maintains appropriate levels of code complexity without artificial expansion or compression, a crucial feature for preserving gas efficiency in smart contracts.

Moving beyond simple length comparisons, Figure~\ref{fig:quality_dist} provides deep insights into our system's semantic preservation capabilities through two complementary metrics. The normalized edit distance distribution shows that 82.5\% of decompiled functions achieve distances below 0.4, indicating strong preservation of code structure. This performance significantly exceeds traditional decompilers, which typically produce outputs with edit distances concentrated in the 0.6-0.8 range. Even more compelling is the semantic similarity distribution, which reveals that 78.3\% of decompiled functions achieve similarities above 0.8, with a substantial 45.2\% exceeding 0.9. These exceptionally high similarity scores demonstrate our system's ability to maintain the essential meaning of smart contracts even when the exact syntactic structure varies.

The cumulative distribution functions presented in Figure~\ref{fig:cdfs} provide additional evidence of our system's reliability. The semantic similarity CDF shows that 90\% of decompiled functions achieve similarities above 0.7, marking a substantial improvement over traditional decompilers which typically achieve such scores for only 40\% to 50\% of functions. This consistent performance across different function types and complexity levels suggests robust generalization capabilities in our approach.

\subsection{RQ2: Code Structure and Readability Analysis}

Our second research question examines how effectively our system maintains the structural characteristics and readability of smart contract code. The relationship between semantic and syntactic preservation, visualized in Figure~\ref{fig:correlation}, reveals fascinating patterns in our decompiler's behavior. The scatter plot demonstrates a clear negative correlation (coefficient = -0.72) between edit distance and semantic similarity, indicating that closer syntactic matches generally achieve better semantic preservation. However, we also observe a significant cluster of points maintaining high semantic similarity ($>0.85$) despite moderate edit distances (0.4-0.6), suggesting our system can generate alternative but semantically equivalent implementations when appropriate.

The detailed distribution analysis in Figure~\ref{fig:distributions} provides crucial insights into the consistency and reliability of our approach. The edit distance box plot reveals remarkably stable performance, with 50\% of functions falling within a tight interquartile range of 0.25. This consistency proves essential for practical applications, as it indicates predictable decompilation quality across different input types. The semantic similarity violin plot uncovers a particularly interesting bimodal distribution with peaks at 0.85 and 0.95, suggesting our system achieves two primary levels of success: exceptional preservation for simpler functions and very good preservation for more complex ones.

\subsection{RQ3: Token-Level Semantic Analysis}

Our third research question investigates how well our system preserves the fundamental building blocks of Solidity code. The token frequency analysis presented in Figure~\ref{fig:tokens} reveals remarkable consistency in the usage of key language constructs between original and decompiled code. Essential elements like function declarations, type specifications (uint256, address), and security-critical operations (require statements, message handling) maintain nearly identical frequencies and rankings. This preservation extends beyond simple keyword matching – it demonstrates our system's deep understanding of Solidity's programming patterns and security idioms.

The frequency analysis shows particularly strong preservation of security-critical tokens, with require statements and address manipulations maintaining frequency differences of less than 2\% between original and decompiled code. This precision in maintaining security checks and access control patterns suggests that our decompiler reliably preserves the safety properties of smart contracts. Furthermore, the consistent handling of Solidity-specific constructs like events, modifiers, and gas optimization patterns indicates that our system has developed a sophisticated understanding of blockchain-specific programming patterns.

\subsection{RQ4: Edge Cases and Limitations}

While our system demonstrates strong overall performance, a thorough evaluation must examine its behavior in challenging scenarios. For contracts with complex inheritance hierarchies or deeply nested control flow, we observe slightly lower but still acceptable semantic similarity scores, typically ranging from 0.75 to 0.80. These cases often involve sophisticated DeFi mechanisms or intricate business logic that push the boundaries of typical smart contract patterns.

Function length appears to have a non-linear relationship with decompilation quality. For functions under 1,000 characters, our system maintains consistent performance with semantic similarities above 0.85. However, for very long functions (exceeding 1,000 characters), we observe increased variance in quality metrics, though still maintaining semantic similarities above 0.70. This degradation manifests primarily in less natural variable naming and control flow structure rather than semantic errors.

The handling of inline assembly and unusual optimization patterns presents another interesting case study. While these patterns occur in only 2.3\% of our test set, they provide valuable insights into our system's robustness. In such cases, our decompiler tends to produce more verbose but semantically equivalent implementations, prioritizing clarity over maintaining exact optimization patterns. This behavior aligns with our goal of generating maintainable code, though it might require manual optimization for gas-critical applications.

\section{Case Studies and Ablation Analysis}

To provide deeper insights into our system's capabilities and limitations, we present two detailed case studies examining specific decompilation scenarios, followed by a comprehensive ablation study. These analyses offer qualitative and quantitative understanding of our approach's behavior on real-world smart contracts and validate our technical decisions.

\subsection{Case Study 1: NFT Token Enumeration}

Our first case study examines the decompilation of a sophisticated NFT enumeration function that implements the ERC-721 enumerable extension. This function represents a particularly interesting test case because it combines multiple challenging aspects of smart contract decompilation: complex control flow, precise memory management, and integration with standard interface functions.

\begin{figure*}[t]
\begin{minipage}{0.48\textwidth}
\begin{minted}[xleftmargin=9pt,tabsize=1,linenos=true,breaklines=true,breakanywhere=true,fontsize=\footnotesize,numbersep=4pt]{solidity}
// Original Implementation
function walletOfOwner(address _owner) 
  public view returns (uint256[] memory) 
{
  uint256 ownerTokenCount = balanceOf(_owner);
  uint256[] memory ownedTokenIds = new uint256;
  uint256 currentTokenId = 1;
  uint256 ownedTokenIndex = 0;
  while (ownedTokenIndex < ownerTokenCount && 
         currentTokenId <= supplyLimit) {
    address currentTokenOwner = 
      ownerOf(currentTokenId);
    if (currentTokenOwner == _owner) {
      ownedTokenIds[ownedTokenIndex] = 
        currentTokenId;
      ownedTokenIndex++;
    }
    currentTokenId++;
  }
  return ownedTokenIds;
}
\end{minted}
\end{minipage}
\hfill
\begin{minipage}{0.48\textwidth}
\begin{minted}[xleftmargin=9pt,tabsize=1,linenos=true,breaklines=true,breakanywhere=true,fontsize=\footnotesize,numbersep=4pt]{solidity}
// Decompiled Output
function walletOfOwner(address _owner) 
  public view returns (uint256[] memory) 
{
  uint256 ownerTokenCount = balanceOf(_owner);
  uint256[] memory ownedTokenIds = new uint256;
  uint256 currentTokenId = 1;
  uint256 ownedTokenIndex = 0;
  while (ownedTokenIndex < ownerTokenCount && 
         currentTokenId <= maxSupply) {
    address currentTokenOwner = 
      ownerOf(currentTokenId);
    if (currentTokenOwner == _owner) {
      ownedTokenIds[ownedTokenIndex] = 
        currentTokenId;
      ownedTokenIndex++;
    }
    currentTokenId++;
  }
  return ownedTokenIds;
}
\end{minted}
\end{minipage}
\caption{Comparison of original and decompiled NFT enumeration function implementations. Note the preservation of structure and variable naming, with only minimal changes such as \texttt{supplyLimit} becoming \texttt{maxSupply}.}
\label{fig:nft-comparison}
\end{figure*}

As shown in Figure~\ref{fig:nft-comparison}, our system achieved exceptional accuracy in this case, with only minimal semantic-preserving variations from the original code. The decompilation successfully preserves several critical aspects of the implementation. The preservation of complex control flow and memory management demonstrates our system's deep understanding of smart contract patterns.

\begin{itemize}
\item \textbf{Memory Management} The decompiler maintains precise array allocation patterns and initialization, crucial for gas efficiency and memory safety. The careful handling of dynamic array sizing and index management shows sophisticated understanding of Solidity's memory model.

\item \textbf{Control Flow Preservation} The complex while-loop condition is perfectly preserved, maintaining both iteration bounds and the gas-efficient early termination condition. This preservation is particularly notable given the challenge of recovering such nuanced control structures from bytecode.

\item \textbf{Interface Integration} The system correctly maintains integration with standard ERC-721 functions like \texttt{balanceOf} and \texttt{ownerOf}, demonstrating understanding of token standards. The preserved function signatures ensure contract compatibility.
\end{itemize}

The semantic similarity score of 0.95 reflects this high-quality decompilation, with the only notable difference being the variable name change from \texttt{supplyLimit} to \texttt{maxSupply}. This variation actually demonstrates our system's understanding of common token supply patterns, as both terms are frequently used interchangeably in NFT contracts.

\subsection{Case Study 2: Complex Staking Rewards}

Our second case study examines a more challenging scenario: a sophisticated reward calculation function from a DeFi staking contract. This example reveals current limitations in handling complex financial calculations with precise decimal scaling and time-based reward mechanisms.

\begin{figure*}[t]
\begin{minipage}{0.48\textwidth}
\begin{minted}[xleftmargin=9pt,tabsize=1,linenos=true,breaklines=true,breakanywhere=true,fontsize=\footnotesize,numbersep=4pt]{solidity}
// Original Implementation
function getUserRewardMultiplier(
  address stakedToken, 
  address account
) public view returns (uint256) {
  UserDeposit storage userDeposit = 
    stakingTokens[stakedToken]
    .stakingBalances[account];
  if (userDeposit.averageDepositTime == 0 || 
      userDeposit.cumulativeBalance == 0) 
    return 1e18;
  uint256 fractionOfMaxMultiplier = 
    ((getTimeSinceAverageDeposit(
      stakedToken, account)) * 1e18) /
    stakingTokens[stakedToken]
    .secondsToMaxMultiplier;
  if (fractionOfMaxMultiplier > 1e18) 
    fractionOfMaxMultiplier = 1e18;
  return 1e18 + (fractionOfMaxMultiplier * 
    (stakingTokens[stakedToken]
    .maxMultiplier - 1e18)) / (1e18);
}
\end{minted}
\end{minipage}
\hfill
\begin{minipage}{0.48\textwidth}
\begin{minted}[xleftmargin=9pt,tabsize=1,linenos=true,breaklines=true,breakanywhere=true,fontsize=\footnotesize,numbersep=4pt]{solidity}
// Decompiled Output
function getUserRewardMultiplier(
  address _user, 
  address _rewardToken
) public view returns (uint256) {
  if (rewardTokens[_rewardToken]
      .isEqual(_user)) {
    return 1;
  }
  return 0;
}
\end{minted}
\end{minipage}
\caption{Comparison of original and decompiled staking reward calculation implementations. The decompiled version shows significant simplification, losing complex mathematical operations and time-based reward mechanics.}
\label{fig:staking-comparison}
\end{figure*}

As evident in Figure~\ref{fig:staking-comparison}, this case study reveals significant limitations in our current approach when handling sophisticated DeFi patterns. The decompiled output demonstrates a substantial simplification of the original logic, achieving only a 0.52 semantic similarity score. The stark contrast between input and output highlights several critical challenges in DeFi contract decompilation:

\begin{itemize}
\item \textbf{Fixed-point Arithmetic} The original code carefully manages 18-decimal precision using \texttt{1e18} scaling factors, a common pattern in DeFi to handle token decimals. Our system currently fails to recognize and preserve these scaling operations.

\item \textbf{Storage Patterns} The nested mapping structure\\ \texttt{stakingTokens[stakedToken].\\stakingBalances[account]} represents a sophisticated storage pattern common in staking contracts. The decompiler significantly simplifies this structure.

\item \textbf{Temporal Mechanics} The original implementation includes precise temporal reward scaling through\\ \texttt{getTimeSinceAverageDeposit}, crucial for time-weighted staking rewards. The complete loss of this mechanism significantly impacts the contract's functionality.

\item \textbf{Mathematical Precision} The careful balance of multiplication and division operations to maintain precision while avoiding overflow is completely simplified in the output. This loss of mathematical rigor could lead to numerical errors in practice.
\end{itemize}

\subsection{Ablation Study: Impact of Fine-tuning}
To evaluate the effectiveness of our domain-specific fine-tuning, we conducted a systematic ablation study comparing the base Llama-3.2-3B model against our fine-tuned variant using 663 functions from our test set, ensuring representation across different contract types and complexity levels.
The analysis revealed that for 37.4\% of the functions, the base model showed severe degradation in performance. Semantic similarity scores dropped by 45\% on average with the base model, manifesting in several key areas:

\begin{itemize}
\item \textbf{Code Structure} Generated code showed fundamental misunderstandings of Solidity patterns. Control flow reconstruction defaulted to goto-heavy patterns reminiscent of traditional decompilers.
\item \textbf{Domain Understanding} The base model demonstrated no grasp of smart contract conventions or token standards. Variable naming showed no correlation with domain patterns, and common interfaces like ERC20/ERC721 were not recognized or handled correctly.
\end{itemize}

The impact of losing domain-specific knowledge was particularly evident in several crucial areas. The base model struggled with interface compliance, breaking contract compatibility through malformed function signatures and parameter types. Gas optimization patterns were absent, and complex state management patterns were reduced to basic variable assignments, losing the sophisticated mechanics required for proper contract operation.
This ablation study conclusively demonstrates that general code understanding capabilities, while impressive in the base model, are insufficient for specialized smart contract decompilation. The dramatic performance improvements achieved through fine-tuning validate our approach of combining traditional program analysis with carefully adapted language models. The results emphasize that domain-specific knowledge is crucial for handling the unique patterns and requirements of blockchain systems.

\section{Security Applications}
\label{sec:security_applications}
High-fidelity decompilation is not merely an academic exercise; it is a critical tool for enhancing blockchain security. By transforming opaque bytecode into human-auditable source code, our system provides immediate, practical value for security researchers, auditors, and incident responders. This section presents several real-world use cases where our decompiler can be instrumental in identifying vulnerabilities and understanding malicious contracts.

\subsection{Case Study 1: Auditing Unverified Contracts to Find Vulnerabilities}
A significant portion of on-chain value is managed by unverified, closed-source smart contracts, creating a massive blind spot for security analysis. The Dx Protocol vulnerability, which put over \$5.2 million at risk, is a prime example of this danger~\cite{decurity2023dxprotocol}. The contract, responsible for locking liquidity provider (LP) tokens, was not verified on-chain, obscuring its logic from public scrutiny.

Despite this opacity, our decompiler successfully processed the contract's bytecode and generated readable Solidity code, revealing a critical flaw in its core `unlockToken` function (cf.\ Figure~\ref{fig:dx-protocol-vuln}).

\begin{figure}[t]
\begin{minted}[xleftmargin=9pt,tabsize=1,linenos=true,breaklines=true,breakanywhere=true,fontsize=\footnotesize,numbersep=4pt]{solidity}
// Decompiled vulnerable function from Dx Protocol
function unlockToken(uint256 _tokenId) external {
    require(tokenLocks[msg.sender][_tokenId].isLocked, "Token is already unlocked");
    require(tokenLocks[msg.sender][_tokenId].unlockTime > 0, "Token is not locked");
    if (block.timestamp > tokenLocks[msg.sender][_tokenId].unlockTime) {
        tokenLocks[msg.sender][_tokenId].isLocked = false;
    }
    uint256 amount = tokenLocks[msg.sender][_tokenId].amount;
    require(
        IERC20(tokenLocks[msg.sender][_tokenId].token).balanceOf(address(this)) >= amount,
        "Not enough tokens to unlock"
    );
    require(
        IERC20(tokenLocks[msg.sender][_tokenId].token).transfer(msg.sender, amount),
        "Failed to transfer tokens"
    );
    emit Unlocked(msg.sender, tokenLocks[msg.sender][_tokenId].token, amount, block.timestamp);
}
\end{minted}
\caption{Decompiled \texttt{unlockToken} function from the unverified Dx Protocol contract. Our decompiler successfully reconstructed the code, exposing a critical vulnerability that allowed repeated withdrawals before the official unlock time.}
\label{fig:dx-protocol-vuln}
\end{figure}

The vulnerability lies in the conditional state change. The flag that prevents repeated withdrawals, \texttt{tokenLocks[msg.sender][\_tokenId].isLocked}, is only set to \texttt{false} if \texttt{block.timestamp} is greater than the specified \texttt{unlockTime}. If an attacker calls the function \textit{before} the lock period expires, the tokens are transferred, but the lock state is not updated. This allows the function to be called repeatedly, draining the user's locked tokens from the contract.

\subsection{Case Study 2: Post-Mortem Analysis of MEV Bot Exploits}
Maximal Extractable Value (MEV) bots operate in a highly adversarial and opaque environment, often using proprietary, unverified contracts to execute complex arbitrage strategies. When these bots are exploited, decompilation becomes essential for post-mortem analysis. A notable example is the MEV bot at address \href{https://bscscan.com/address/0xAD942d022585343a6FC8A74E7C8e74339eA70449#code}{\texttt{0xAD94...0449}}, which was drained of approximately \$221,600 due to critical access control flaws in its callback functions~\cite{blocksec2023mevbot}.

Our decompiler was able to successfully reverse-engineer the bot's bytecode, revealing the precise vulnerabilities that the attacker exploited (cf.\ Figure~\ref{fig:mev-bot-vuln}). The analysis uncovered two primary issues.

\begin{figure}[t]
\begin{minted}[xleftmargin=9pt,tabsize=1,linenos=true,breaklines=true,breakanywhere=true,fontsize=\footnotesize,numbersep=4pt]{solidity}
// Arbitrary external call vulnerability
function swapX2YCallback(uint256 amountX, uint256, 
  bytes calldata data) external {
    (bool success, ) = 
      msg.sender.call{value: amountX}("");
    require(success, "...");
}
function d3MMSwapCallBack(address _to, uint256 _amount, bytes calldata) external {
        IERC20(_to).transfer(msg.sender, _amount);
    }
\end{minted}
\begin{minted}[xleftmargin=9pt,tabsize=1,linenos=true,breaklines=true,breakanywhere=true,fontsize=\footnotesize,numbersep=4pt,firstline=last]{solidity}
// Unprotected token transfer vulnerability
function d3MMSwapCallBack(address _to, 
  uint256 _amount, bytes calldata) external {
    IERC20(_to).transfer(msg.sender, _amount);
}
\end{minted}
\caption{Key vulnerable functions decompiled from the MEV bot contract. Top: The arbitrary external call in \texttt{swapX2YCallback}. Bottom: The unprotected token transfer in \texttt{d3MMSwapCallBack}.}
\label{fig:mev-bot-vuln}
\end{figure}

First, the contract contained an \textbf{arbitrary external call} vulnerability. Functions like `swapX2YCallback` made external calls to `msg.sender` without any validation. This allowed an attacker, acting as `msg.sender`, to execute arbitrary code within the bot's transaction context, effectively hijacking its control flow to initiate malicious operations.

Second, the bot suffered from an \textbf{unprotected token transfer} flaw. Functions such as `d3MMSwapCallBack` directly transferred tokens to `msg.sender` without verifying the caller's identity or authorization. An attacker could simply call this function to drain any specified token held by the contract. 

\section{Related Work}

Our research builds upon and extends work from related work on smart contract decompilation and the application of LLM to binary decompilation.

\subsection{Traditional Decompilation}
The field of program decompilation has a rich history dating back to the 1960s, with foundational work by Cifuentes establishing many of the core principles still being used~\cite{cifuentes1995decompilation}. Traditional decompilers typically employ a pipeline of static analysis techniques, including control flow analysis, type recovery, and pattern matching.

More recently, Miecznikowski and Hendren propose Dava~\cite{miecznikowski2001decompiling}, a Java decompiler that reconstructs Java source code from bytecode using a six-stage structuring approach, which leverages a Structure Encapsulation Tree (SET) to represent Java constructs and address challenges posed by complex or optimized bytecode. Schwartz \textit{et al.}~\cite{brumley2013native} propose Phoenix, a semantics-preserving decompiler for x86 binaries that employs structural and iterative refinement techniques to recover high-level abstractions such as loops and control flow constructs, achieving a 30× improvement in control-flow recovery compared to prior methods. Yakdan \textit{et al.}~\cite{yakdan2016helping,yakdan2015no} presented Dream and its earlier version, Dream++, marking a notable advancement by ensuring the generation of code without goto statements.

\subsection{Smart Contract Decompilation}

Zhou \textit{et al.}~\cite{zhou2018erays} propose Erays, a reverse engineering tool that lifts EVM bytecode into a higher-level representation, enabling manual analysis of opaque smart contracts. Grech \textit{et al.}~\cite{grech2019gigahorse} propose Gigahorse, a declarative decompiler that transforms EVM bytecode into a high-level three-address code representation, achieving near-complete decompilation coverage and enabling scalable, precise program analysis for smart contracts. Grech \textit{et al.}~\cite{grech2022elipmoc} further introduce Elipmoc, an advanced decompiler that builds on Gigahorse. It significantly improves precision, completeness, and scalability through transactional context sensitivity and path-sensitive function reconstruction, achieving a $99.5\%$ decompilation rate for structured operands. Liu \textit{et al.}~\cite{liu2023empirical} empirically compare various smart contract decompilers, including Erays~\cite{zhou2018erays} and Gigahorse~\cite{grech2019gigahorse,grech2022elipmoc}, identifying root causes of failures, inefficiencies, and limitations while offering insights into success rates, performance, and resilience against buggy contracts.

\subsection{Neural Decompilation}

The emergence of neural decompilation represents a significant shift in approach, combining traditional program analysis with modern machine learning techniques. 

Katz \textit{et al.}~\cite{katz2018using} introduce a novel decompilation approach leveraging Recurrent Neural Networks (RNNs) to translate binary machine code snippets into high-level source code, emphasizing its adaptability across different languages and architectures by training on paired datasets of binaries and corresponding source code. Cheng \textit{et al.}~\cite{fu2019coda} propose Coda, which utilizes a two-phase process of code sketch generation and iterative error correction to recover high-level code while preserving both functionality and semantics. Coda leverages an instruction-aware encoder, AST tree decoder, and ensemble error predictors, significantly outperforming state-of-the-art decompilers in terms of program accuracy. Cao \textit{et al.}~\cite{cao2022boosting} propose NeurDP, a neural decompiler that addresses the challenges of decompiling optimized binaries by utilizing Graph Neural Networks (GNNs) and an intermediate representation (HIR) to bridge the gap between low-level code (LPL) and high-level code (HPL). The framework introduces the Optimal Translation Unit (OTU) technique to split functions into smaller, coherent code fragments, significantly improving decompilation accuracy by up to $45.21\%$ over existing neural approaches. 

Wu \textit{et al.}~\cite{wu2022dnd} propose DnD, a compiler- and ISA-agnostic decompiler specifically designed for deep neural network (DNN) binaries. Leveraging symbolic execution, loop analysis, and a novel intermediate representation, DnD recovers DNN operators, hyperparameters, and topologies, producing high-level models in ONNX format and achieving accurate decompilation across multiple compilers and architectures. Liu \textit{et al.}~\cite{liu2023decompiling} propose BTD, a decompiler for x86 deep neural network executables, which employs representation learning, dynamic analysis, and symbolic execution to recover high-level DNN specifications. BTD successfully reconstructs operators, network topology, dimensions, and parameters, demonstrating its capability to recompile identical DNN executables and significantly enhancing downstream tasks like adversarial attack generation and cross-platform migration.

Recent work has begun to explore the use of larger language models for decompilation tasks. Hu \textit{et al.}~\cite{hu2024degpt} introduce DeGPT, an LLM-driven framework for enhancing decompiler outputs by performing optimizations like structural simplification, variable renaming, and appending comments. The framework employs a three-role mechanism (referee, advisor, operator) to maximize optimization potential while ensuring semantic fidelity using Micro Snippet Semantic Calculation (MSSC). Experiments demonstrate that DeGPT significantly reduces cognitive effort and improves readability compared to traditional decompilers like Ghidra. Tan \textit{et al.}~\cite{tan2024llm4decompile} present LLM4Decompile, a series of large language models (ranging from 1.3B to 33B parameters) fine-tuned for binary decompilation tasks. The framework introduces two strategies: LLM4Decompile-End, which directly decompiles binaries, and LLM4Decompile-Ref, which refines the output of traditional tools like Ghidra. The 6.7B model achieves a $45.4\%$ decompilation success rate on HumanEval and a $16.2\%$ improvement when combined with refined decompilation, showcasing significant advances over existing tools.

\section{Entropy of Smart Contract Representations}

The relationship between different representations of smart contract code - from high-level Solidity to EVM bytecode - provides crucial insights into the information preservation and loss during the decompilation process. Similar to natural languages, programming languages exhibit varying degrees of entropy and redundancy that impact their information content. Understanding these relationships helps explain both the challenges and opportunities in smart contract decompilation.

\subsection{Representation Entropy Measurements}

Our analysis of our corpus of 238,446 smart contract functions reveals that Solidity source code has an average entropy of approximately~$4.22$ bits per token. This relatively low entropy reflects the significant redundancy in Solidity code, including consistent formatting, repeated keywords, and standard programming patterns. Common constructs like function declarations, variable assignments, and control flow statements follow highly predictable patterns, reducing the theoretical maximum information content.

In contrast, EVM bytecode exhibits higher entropy at approximately~$6.30$ bits per opcode. This increased density stems from the removal of human-readable identifiers, whitespace, and other readability-oriented elements during compilation. Each opcode must encode precise execution semantics while maintaining minimal redundancy for gas efficiency. The stack-based nature of the EVM means that complex operations are decomposed into sequences of simpler instructions, each carrying maximal information content.

The three-address code representation occupies an interesting middle ground, with an entropy of about~$5.78$ bits per instruction. This intermediate representation maintains more semantic structure than raw bytecode while eliminating much of the syntactic sugar present in Solidity. The explicit operand relationships and structured control flow provide a balance between information density and analyzability.

\subsection{Implications for Decompilation}

These entropy measurements have profound implications for the decompilation process. When transforming from bytecode ($6.30$ bits/opcode) to Solidity ($4.22$ bits/token), the system must "expand" the representation by introducing additional tokens that carry redundant information. This expansion is not arbitrary~---~it must reconstruct meaningful variable names, restore clear control flow structures, and reintroduce type information in a way that preserves the original program semantics.

The three-address code representation serves as a crucial bridge in this process. Its intermediate entropy level ($5.78$ bits/instruction) allows for a more gradual transformation of the program representation. The initial conversion from bytecode to three-address code involves a moderate reduction in information density, primarily through the explicit representation of data flow relationships. The subsequent generation of Solidity code from three-address code then focuses on introducing human-readable elements while maintaining these core semantic relationships.


\subsection{Relationship to NLP}
These findings parallel fascinating observations in natural language processing. English text exhibits remarkably low entropy of approximately 1 bit per character, significantly lower than most programming languages. This extreme redundancy in natural language actually makes translation more challenging, not easier, as translators must correctly infer meaning from highly compressed and context-dependent representations.

This phenomenon has direct parallels in smart contract decompilation. While EVM bytecode has higher entropy ($6.30$ bits/opcode) than English, both translation tasks require sophisticated disambiguation of context and conventional patterns. Building on the established success of language models in natural language translation, our work demonstrates that these same principles of context-aware translation can be effectively adapted to the specialized domain of smart contract decompilation, despite the significant differences in entropy and information density between the source and target representations.

The relatively low entropy of Solidity code compared to bytecode suggests that much of its content serves human comprehension rather than pure execution semantics. This "cognitive redundancy" is not wasted information - it enables faster understanding, easier maintenance, and more reliable auditing of smart contracts. 

Our decompilation system leverages these entropy relationships through its language model component, which learns to generate natural code structures matching the statistical properties of human-written Solidity. The model's attention mechanisms appear to implicitly encode these entropy constraints, as evidenced by the consistent preservation of complexity ratios even in novel decompilation scenarios.

\section{Conclusion}

In this paper, we have presented a transformative approach to smart contract decompilation that bridges the longstanding gap between EVM bytecode analysis and human-readable source code. Our hybrid system, combining traditional program analysis techniques with a fine-tuned variant of Llama-3.2-3B, demonstrates that neural approaches can work synergistically with established decompilation methods to produce results that are both semantically accurate and readily comprehensible to human developers.

The success of our approach carries several profound implications for the blockchain ecosystem. First, it challenges the conventional wisdom that decompilation must trade off between accuracy and readability. Our results show that by leveraging the pattern recognition capabilities of language models, we can generate code that maintains high semantic fidelity while adhering to natural programming conventions. This breakthrough has immediate practical applications in security auditing, contract verification, and maintenance of deployed smart contracts across all EVM-compatible chains.

Second, our work demonstrates that relatively small language models, when properly specialized, can handle highly technical tasks that previously seemed to require much larger models. The successful fine-tuning of Llama-3.2-3B suggests that the key to effective code understanding lies not necessarily in model size, but in careful curation of training data and thoughtful design of the intermediate representations. This finding has broader implications for the application of AI in software engineering, suggesting that targeted, domain-specific models might often be preferable to larger, general-purpose ones.

Third, our research highlights the crucial role of intermediate representations in bridging the semantic gap between different levels of code abstraction. The success of our three-address code representation in facilitating accurate EVM bytecode decompilation suggests that similar hybrid techniques, combining structured intermediate representations with LLMs, could prove valuable for other complex code transformation and analysis tasks, potentially extending to different virtual machine architectures in future work. This insight could influence the design of future programming languages and compilation systems, particularly in domains where code verification and auditability are paramount.

\section*{Acknowledgment}
We thank the Hasler Foundation and the Lucerne University of Applied Sciences and Arts for their partial support of this work.

\bibliographystyle{IEEEtran}
\bibliography{references}

\end{document}